# Effectiveness of Multi-Layered Radiation Shields Constructed from Polyethylene and Metal Hydrides Using HZETRN and OLTARIS for space applications

[1]Sreedevi V V, [2*]Kavita Lalwani

Department of Physics, Malaviya National Institute of Technology Jaipur, India

Corresponding author: *kavita.phy@mnit.ac.in

**Abstract**

A major challenge for extended human spaceflights in deep space is the dangerous exposure to space radiation. In previous studies aluminium has been used as a multilayer shielding material in GCR space radiation environment with high dose equivalent. To further reduce the dose equivalent, shielding effectiveness of various metal hydrides in GCR free space environment is investigated using HZETRN2015 (High charge (Z) & Energy TRaNsport) and OLTARIS (On-Line Tool for the Assessment of Radiation In Space) in this work. Metal hydride materials are chosen because of their capacity to store hydrogen. Among these materials, lithium hydride has demonstrated superior effectiveness as a radiation shield. Given this, the potential of a multilayer shield composed of polyethylene and lithium hydride is also being explored considering the tensile strength certain varieties of polyethylene like UHMWPE (Ultra High Molecular Weight Poly-Ethylene) can provide. The results from HZETRN2015 and OLTARIS transport codes are compared and found in agreement.

**Key Words:** Space Radiation, GCR, HZETRN, OLTARIS, Hydrides, Polyethylene, LiH.

## 1. Introduction

One of the main challenges for prolonged human space expeditions is space radiation. Unlike on earth, the interplanetary space is constantly being bombarded with ionising radiation [1]. Exposure to such high radiation for long can do severe harm to the human body as well as electronic equipment. Space radiation cannot reach earth's surface because of the strong magnetic field and thick atmosphere of earth. There are three main sources for space radiation: Galactic Cosmic Rays (GCRs), Solar Particle Events (SPEs) and Trapped particles. The GCRs originate from energetic sources outside the solar system such as stars and supernovae and hence they are isotropic in terms of direction. They consist of 85% protons and 13% of alpha particles followed by completely ionised nuclei of elements with higher atomic number [1]. Solar Energetic Particles (SEPs) [2] are released through energetic processes in the solar corona like solar flares and Coronal Mass Ejections (CMEs). They consist mainly of protons (90-95%), followed by helium nuclei (5-8%), along with some heavier nuclei (1%). These compositions differ between various events. SPEs are in general unpredictable. Mostly they occur at times of maximum solar activity. The Earth's magnetosphere captures high-energy radiation particles, protecting the planet from solar storms and the continuous flow of solar wind. These trapped particles form two radiation belts, known as the Van Allen Belts, which encircle the Earth . They can pose a threat to satellites, which are orbiting in Low Earth Orbits (LEO). Space radiation presents significant health hazards, including an elevated risk of cancer, acute radiation sickness, and possible harm to the nervous system. Astronauts subjected to high levels of radiation may also face cognitive challenges and an increased likelihood of cardiovascular diseases. For long-duration missions like Mars exploration, ensuring the health and safety of astronauts is crucial and cannot be compromised. This is why innovations in space radiation shielding technology are essential.

Two main approaches for space radiation shielding are active and passive shielding methods. Active shielding uses electromagnetic fields to divert harmful radiation. This advanced technique lowers direct exposure, but necessitates energy input and intricate systems to sustain its effectiveness [3]. In passive shielding, a physical material is employed, which can effectively attenuate the incoming radiation. Heavy charged particles lose their energy in shielding materials through electronic and nuclear fragmentation. This study primarily focuses on passive shielding techniques.

## 2. Materials and Simulation Setup

### 2.1 Radiation Shielding Materials:

The material used for shielding in a spacecraft should have high shielding efficiency as well as some additional properties such as high tensile strength and low reactivity. Generally, aluminium is used in spacecrafts due to light weight and structural integrity. But many other materials are found to have produced less absorbed dose than aluminium. As far, hydrogen is found to be the most effective shielding material in terms of dose reduction. But there are practical difficulties in using hydrogen as a shielding material in spacecraft as it is highly unstable. Hydrogen containing compounds also exhibits high shielding effectiveness. Polyethylene, being the hydrocarbon chain with highest hydrogen content, is found to be more effective than many other polymers and conventional materials like aluminium. As a polymer, its structural properties can be modified. UHMWPE (Ultra High Molecular Weight Poly-Ethylene) [4] is a specialised form of polyethylene, characterised as a linear, semi-crystalline homopolymer with an exceptionally high molecular mass. UHMWPE fibres offer excellent radiation shielding, high tensile strength, low density, superior impact resistance, and durability in space environments, making them a promising material for future space structure applications. Metal hydrides are known for their hydrogen storing capacity [5]. This paper explores the shielding effectiveness of various metal hydrides [6,7], as well as multilayer shields [8] combining polyethylene and metal hydrides. The results from two numerical simulations, HZETRN 2015 [9] and OLTARIS [10] are compared. The list of optimized shielding materials are listed in table 1.

| S.No. | Material Name | Chemical Formula | Density in g/cm$^3$ |
|---|---|---|---|
| 1 | Aluminium | Al | 2.7 |
| 2 | Polyethylene | $C_2H_4$ | 0.96 |
| 3 | Beryllium Borohydride | Be$(BH_4)_2$ | 0.604 |
| 4 | Ammonia Borane | $NH_3BH_3$ | 0.78 |
| 5 | Super Hydride | $Li(C_2H_5)_3BH$ | 0.89 |

| 6 | Beryllium Hydride | $BeH_2$ | 0.65 |
| 7 | Lithium Borohydride | $LiBH_4$ | 0.68 |
| 8 | Lithium Hydride | $LiH$ | 0.82 |

Table 1: List of Optimised Shielding Materials.

## 2.2 HZETRN 2015

The HZETRN (High charge (Z) and Energy TRaNsport) code [9] is a deterministic transport model designed by NASA specifically for simulating space radiation transport. The HZETRN code employs numerical solutions to the time-independent, linear Boltzmann equation, utilizing the continuous slowing down approximation. In this approach, discrete atomic interactions are modeled through stopping power. The code has undergone regular updates, with notable releases in 2010, 2015, and 2020, named HZETRN2010, HZETRN2015, and HZETRN2020, respectively. HZETRN2015 was employed for all calculations in this study. It features algorithms and options for 3D transport simulations within user-defined combinatorial or ray-trace geometries. Additionally, it includes computationally efficient bi-directional transport algorithms, similar to those in HZETRN2010, for handling multilayer slab transport. Users can also generate an interpolation database for varying thicknesses across one to three user-defined material layers using a straight-ahead transport algorithm, like that in HZETRN2005.

The code supports transport calculations for Galactic Cosmic Rays (GCR), Solar Particle Events (SPE), Low Earth Orbit (LEO), and custom environmental boundary conditions. For SPE and LEO scenarios, it accounts for the transport of neutrons, protons, and light ions, while GCR boundary conditions extend to include heavy ions, pions, muons, electrons, positrons, and photons.

## 2.2 OLTARIS

OLTARIS (On-Line Tool for the Assessment of Radiation In Space) [10] is a web-based platform that leverages HZETRN to enable scientists and engineers to analyze the impacts of space radiation on humans and electronic systems. The OLTARIS architecture consists of two primary components: the website, where users interact through a browser, and the execution environment, which handles the computations. The website is primarily built using standard open-source components, while the execution environment relies on FORTRAN executables running on a computational cluster. A flowchart illustrating the data and execution flow is shown in figure 1 [11].

The flowchart uses different box colors to represent various system elements: green boxes indicate user-supplied data, blue boxes represent data that can be either downloaded from the web server or used in calculations and stored on the execution host, and gold boxes denote computations performed on the execution host. This modular design facilitates maintenance and upgrades as new algorithms, methods, and features are developed.

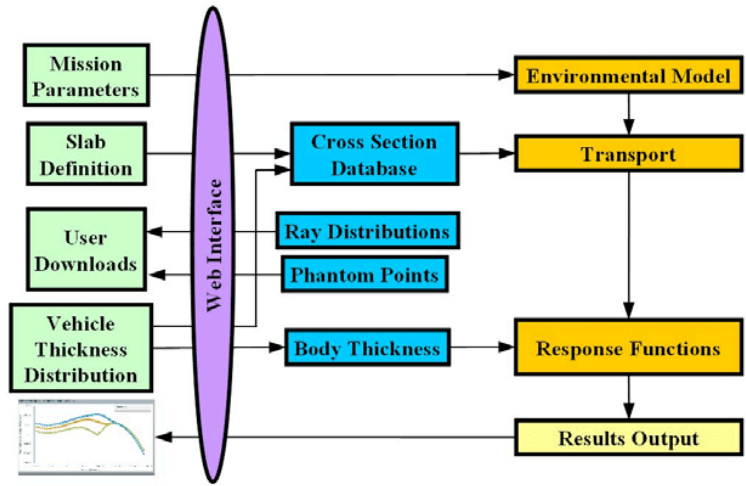

Figure 1: OLTARIS data and execution flowchart [11].

## 3. Results

In the following subsections, various studies involving dose equivalent of shielding materials are discussed systematically. All the studies, which are presented are executed in GCR free space environment using the Badhwar O'Neill 2014 model. The solar modulation parameter, φ = 475 MV, is chosen to represent solar minimum conditions for a mission of duration 1 day. The response is calculated in tissue with quality factor ICRP 60 [12]. These parameters are identical in both HZETRN and OLTARIS.

### 3.1. Dose Equivalent for variable thickness of shielding materials

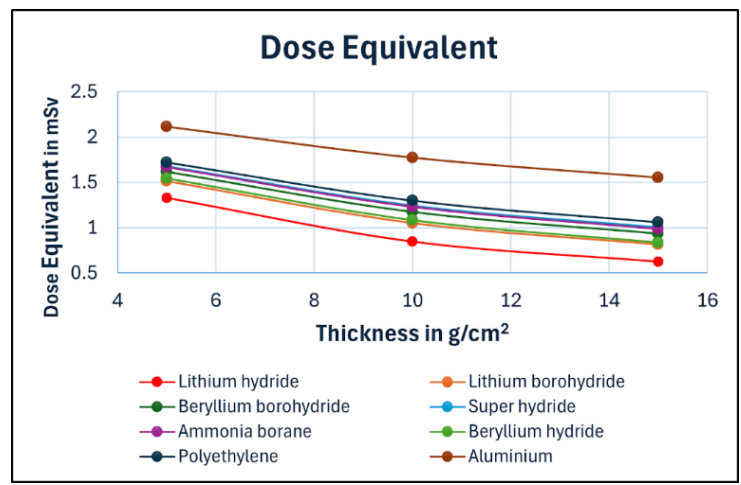

Figure 2: Variation of dose equivalent with thickness of various hydrides, polyethylene and aluminium.

The dose equivalent of different materials is computed by varying their thickness from 5 to 15 g/cm² in HZETRN 2015 [9]. Spherical geometry is used to run the simulation. The results are shown in figure 2. Beyond 15 g/cm², there is no considerable reduction in dose. The dose equivalent of metal hydrides is less than polyethylene and aluminium. Lithium hydride offers much lesser dose equivalent than other hydrides.

### 3.2. Dose Equivalent by Different Particles

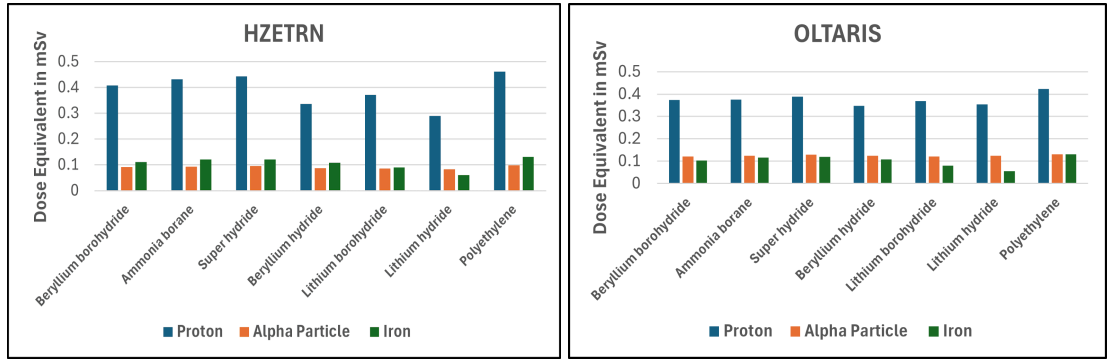

Figure 3: Particle wise dose equivalent of metal hydrides and polyethylene using HZETRN (left) and OLTARIS (right).

The particle wise dose equivalent is studied for different incoming particles such as proton, alpha particle and heavy ion (iron) in the GCR spectrum. Proton and alpha particle are the particles with high relative abundance in the GCR spectrum and iron represents the HZE (High Charge and Energy) particle category. The results are shown in figure 3. The highest dose equivalent is due to proton, which is clearly due to its abundance. The dose equivalent due to iron, despite its low abundance, is comparable to that of alpha particle. The comparison between the metal hydrides aligns with the previous trend. Lithium hydride is effective in reducing proton as well as heavy ion doses. The results from both HZETRN and OLTARIS confirms the relative effectiveness between the different materials. The dose equivalent of proton in HZETRN is in general slightly higher than that in OLTARIS and the dose equivalent of alpha particle is higher in OLTARIS, when compared to that in HZETRN.

### 3.3. Comparison of dose equivalent in HZETRN and OLTARIS

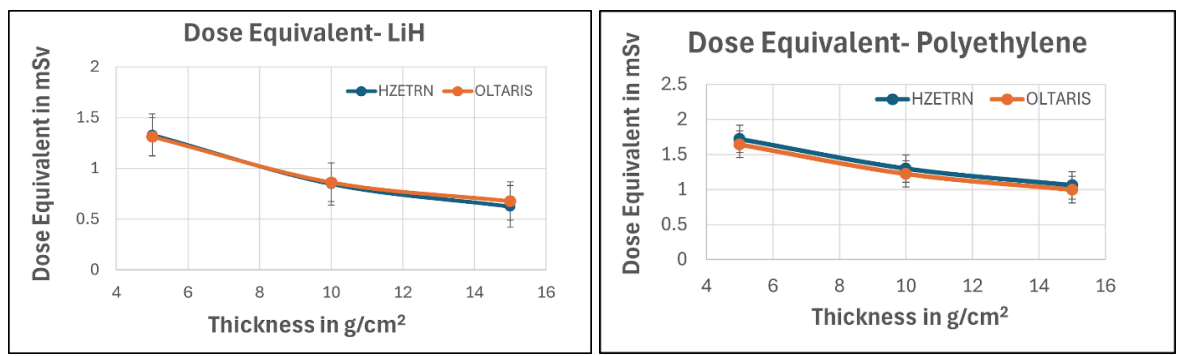

Figure 4: Dose equivalent of lithium hydride (left) and polyethylene (right) - comparison between OLTARIS and HZETRN.

From section 3.1 it can be seen that out of all metal hydrides, LiH proves to be a better shielding material, as it produces the least dose equivalent. Polyethylene is also taken into consideration due to its high tensile strength. In this section we compare the dose equivalent for variable thickness of LiH and polyethylene from HZETRN and OLTARIS. The results are shown in figure 4. For LiH, dose equivalent values from both transport codes are in agreement. For polyethylene, the dose equivalent values from HZETRN is higher than that from OLTARIS by 0.0715 mSv on average.

### 3.4. Comparison of radiation flux in HZETRN and OLTARIS

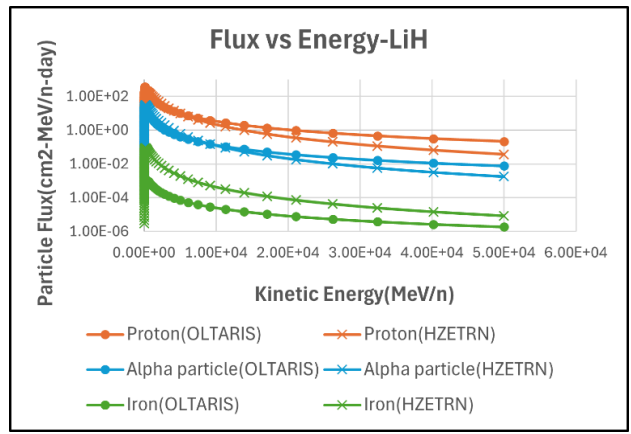

Figure 5: Flux vs energy of 15 g/cm$^2$ lithium hydride using HZETRN and OLTARIS.

Radiation flux after transport as a function of kinetic energy is studied for proton, alpha particle and iron in LiH. The results are shown in figure 5. The relative values of particle flux for proton, alpha particle and iron corresponds to their abundance in the GCR spectrum. At higher energies, the particle flux for each particle reduces. Flux after transport obtained from OLTARIS is higher for proton and alpha particle and is lesser for iron, when compared to the values from HZETRN.

### 3.5. Comparison of Dose Equivalent for Multilayer Shields

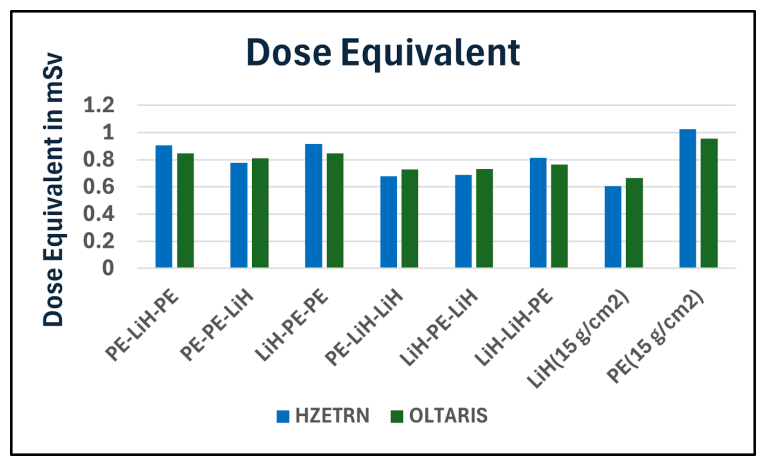

Figure 6: Dose equivalent of different combinations of a multilayer shield using polyethylene and lithium hydride.

A multilayer semi-infinite slab of 15 g/cm$^2$, which consists of three layers of 5 g/cm$^2$ each is created using polyethylene and lithium hydride. Dose equivalent is calculated for all possible combinations of such a shield and is compared with that of 15 g/cm$^2$ LiH slab and 15 g/cm$^2$ polyethylene slab. The results of dose equivalent from OLTARIS and HZETRN are compared and shown in figure 6. The results show that the dose rate not only depends on the material composition of the shield, but also on the relative position of the materials. LiH seems to be more effective, when placed in the innermost layer as compared to when it is placed in the middle or outer layer. The dose equivalent values from HZETRN and OLTARIS are in agreement. Average difference between both sets of results is 0.0537mSv. It is noteworthy that there is only little difference in dose equivalent between 15 g/cm$^2$ LiH slab and PE-LiH-LiH combination.

### 3.6. Variation of Dose Equivalent with depth inside the multilayer shields

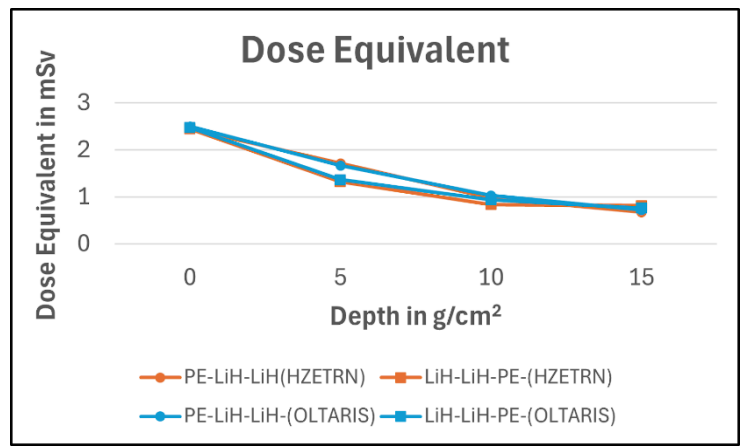

Figure 7: Dose equivalent vs depth for PE-LiH-LiH and LiH-LiH-PE combinations using OLTARIS and HZETRN.

From section 3.5 it is observed that dose equivalent depends on the order of different shielding layers inside the radiation shield. Therefore, it is necessary to investigate the variation in dose equivalent with depth inside the shield. Let us take the cases of two shields with the same material composition: PE-LiH-LiH and LiH-LiH-PE (two LiH layers and one PE layer). The variation of dose equivalent with depth of these two shields is computed and shown in figure 7. It can be seen that maximum dose reduction is achieved in the outermost layer. Polyethylene on the inner layer does not contribute to the dose reduction as when it is placed on the outer layer. The results from both the sources align with this observation.

### 4. Summary

We have investigated the shielding effectiveness of some hydride compounds considering their hydrogen storing capacity. Variation of dose equivalent with the thickness of spherical shield constructed from the hydrides was analysed. Particle wise dose equivalent is also evaluated for all the shielding materials. All the materials performed better than aluminium and polyethylene, among them the best being lithium hydride. Particle flux after transport of 15 g/cm$^2$ LiH was studied. The LiH was combined with polyethylene to construct a multilayer shield, considering the tensile strength polyethylene can provide. Dose equivalent of different combinations of such a shield was calculated. It was found that introducing one layer of polyethylene as an outer layer doesn't increase the dose equivalent of LiH greatly. The results from HZETRN2015 and OLTARIS were compared for each study. In all the cases both results agree with minor differences. The greatest discrepancies in dose equivalent values are approximately 0.07mSv.

## 5. Acknowledgements

We thank MNIT Jaipur for supporting this research work. We are also thankful for OLTARIS and HZETRN transport codes provided by NASA.